\documentclass[prd,11pt,reprint,nofootinbib,superscriptaddress]{revtex4-2}
\usepackage[utf8]{inputenc}
\usepackage[english]{babel}
\usepackage{graphicx}   
\usepackage{dcolumn}    
\usepackage{bm} 
\usepackage{amsmath}    
\usepackage{verbatim}   
\usepackage{color}      
\usepackage{hyperref}
\usepackage{physics}
\usepackage{amssymb}
\usepackage{float}
\usepackage{booktabs}
\usepackage{bigints}

\begin{document}

\title{Emergent Universe Model from Modified Heisenberg Algebra}
\author{G. Barca}\email{gabriele.barca@uniroma1.it}
\affiliation{Department of Physics, “La Sapienza” University of Rome, P.le Aldo Moro 5, 00185 Rome, Italy}\affiliation{INFN, Sezione di Roma 1, P.le A. Moro 2, 00185 Rome, Italy}
\author{G. Montani}\email{giovanni.montani@enea.it}
\affiliation{ENEA, Fusion and Nuclear Safety Department, C. R. Frascati,\\Via E. Fermi 45, 00044 Frascati (RM), Italy}\affiliation{Department of Physics, “La Sapienza” University of Rome, P.le Aldo Moro 5, 00185 Rome, Italy}
\author{A. Melchiorri}\email{alessandro.melchiorri@roma1.infn.it}
\affiliation{INFN, Sezione di Roma 1, P.le A. Moro 2, 00185 Rome, Italy}\affiliation{Department of Physics, “La Sapienza” University of Rome, P.le Aldo Moro 5, 00185 Rome, Italy}

\begin{abstract}
We provide an Emergent Universe picture in which the fine-tuning on the initial conditions is replaced by cut-off physics, implemented on a semiclassical level when referred to the Universe dynamics and on a purely quantum level for the quantum fluctuations of the inflaton field. The adopted cut-off physics is inspired by Polymer Quantum Mechanics but expanded in the limit of a small lattice step. On a quasi-classical level, this results in modified Poisson Brackets for the Hamiltonian Universe dynamics similar to a Generalized Uncertainty Principle algebra. The resulting Universe is indeed asymptotically Einstein-static, emerging from a finite volume configuration in the distant past and then properly reconnecting with the most relevant Universe phases. The calculation of the modifications of the primordial inflaton spectrum is then performed by treating new physics as a small correction on the standard Hamiltonian of each Fourier mode of the field.

The merit of this study is to provide a new paradigm for a non-singular Emergent Universe, which is associated with a precise fingerprint on the temperature distribution of the microwave background, in principle observable by future experiments.
\end{abstract}

\maketitle

\section{Introduction}
One of the most relevant open questions in Relativistic Cosmology concerns the existence of the initial singularity \cite{MontaniPrimordial,Weinberg,KolbTurner,KhalatnikovLifshitz63,MisnerMixmaster68,BKL82}. Indeed, as shown in well-known papers \cite{SingularityTheorems1,SingularityTheorems2}, the existence of a singular instant in the past of our Universe where the curvature invariant diverges and the Einstein equations are no longer predictive is a general feature of the cosmological problem, which has nothing to do with the highly symmetric nature of the Robertson-Walker (RW) geometry describing the isotropic Universe \cite{Weinberg}. For this reason, any physics possibly able to overcome the singularity of the primordial Universe acquires a particular relevance. If the canonical quantization in the Wheeler-DeWitt formulation has been unable to provide a non-singular quantum cosmology \cite{Isham75,Benini2007} (see also \cite{Giovannetti2022} for a different perspective on this scenario), the reformulation in terms of Ashtekar variables in the so-called Loop Quantum Cosmology (LQC) \cite{Ashtekar2011Review} has determined the existence of a Big Bounce, i.e. a Universe with a non-zero minimal volume where the collapsing and the expanding branches of the dynamics are connected and the singularity is avoided (for a review of LQC and Polymer Quantum Mechanics (PQM) \cite{CorichiPQM} approaches to the emergence of a bouncing cosmology, see \cite{Review}). However, a Big Bounce can appear also in classical modified gravity, as discussed for instance in \cite{ImmirziBounce,NiehYanR2Bounce,NiehYanBianchiIBounce}. 

Here, we will consider not a bouncing cosmology, but simply a non-singular cosmology that comes from assigning specific initial conditions on the closed RW model dynamics, known as the ``Emergent Universe'' (EU) \cite{OriginalEU,EU2}. The interest for such an EU model was recently renewed by the analyses of the Planck data sets \cite{Planck2018CMB,Planck2018Cosmo}, which seem to allow for a present-day positive curvature of the Universe \cite{MelchiorriNature,HandleyClosed}.

The possibility to classically solve the singularity is an interesting subject, but for the Emergent Universe this result is valid only for a specific fine-tuning of the initial conditions on the cosmological dynamics. Here we overcome this shortcoming of the original idea by considering suitable modified Poisson brackets inspired by cut-off physics such as PQM \cite{CorichiPQM,Battisti} and the Generalised Uncertainty Principle representation (GUP) \cite{Maggiore,Kempf,BarcaGUP,SebastianoGUP,BossioGUP,FadelMaggiore}, which induce an Emergent Universe scenario still on the classical level, valid for any assignment of the Cauchy problem. After reviewing the original literature on the subject of the classical EU model, we show how it can be obtained thanks to a modified Uncertainty Principle coming from an expanded Polymer formulation for a small enough lattice parameter \cite{BarcaGUP}, when we consider the classical cosmological dynamics via correspondingly modified Poisson algebra. The relevance of this formulation of the EU picture relies on the generality of its non-singular behavior, without the need for a constraint on the initial conditions to be required ab initio. In other words, including a quasi-classical modification of the symplectic algebra similar in its phenomenology to a modified gravity approach, we are able to get an EU with an asymptotic non-singular beginning for the synchronous time approaching negative infinity. We also properly characterize the different phases of the Universe evolution, starting with a radiation-dominated era close to the classical singularity, passing through an inflationary de Sitter period obtained including a constant energy density term, and ending again with a radiation-dominated Universe (the study of a late-time dark energy-dominated era, possible for an EU as mentioned in \cite{OriginalEU}, is beyond the scope of this work). 

An important part of the present analysis is dedicated to the calculation of the primordial Spectrum corrections when the inflaton field obeys the same symplectic algebra at the ground of the obtained quasi-classical dynamics, but implemented on the pure quantum sector. We treat the additional term emerging in the Fourier-decomposed Hamiltonian for the Mukhanov-Sasaki variable, which parameterizes the scalar perturbations \cite{Mukhanov}, as a small perturbation and we determine the corrections it induces on the standard states (associated to a time-dependent harmonic oscillator). As a result, we are able to determine the Spectrum corrections due to the new physics at the ground of our study, and we show under which constraints on the model parameters and initial conditions the modification is a reliably small and potentially observable feature.

We conclude by stressing that the present analysis offers an interesting new perspective on the origin of a non-singular isotropic Universe, whose underlying cut-off physics can leave a precise fingerprint on the profile of the microwave background temperature distribution.

The manuscript is organized as follows. In Section \ref{ham} we present the standard EU in its Hamiltonian formulation. In Section \ref{EUpoly} we introduce the modified algebra and use it to derive a non-fine-tuned EU model; then, in Section \ref{power} we derive the modified Power Spectrum of perturbations using the same modified algebra implemented on the Mukhanov-Sasaki variable. In Section \ref{concl} we conclude the paper with a brief summary and outlook.

\section{Hamiltonian Formulation of the Emergent Universe Model}
\label{ham}
Here we compactly present the standard Emergent Universe model \cite{OriginalEU,EU2} starting from the Hamiltonian formulation of the FLRW homogeneous and isotropic model. We derive a non-singular, ever-expanding solution and then show the potential used to end the inflationary expansion.

\subsection{The Standard Emergent Universe Scenario}
The configurational variables that we will use for the gravitational sector are the volume $v=a^3$, where ${a=a(t)}$ is the dimensionless cosmic scale factor, and its conjugate momentum $p_v\propto\dot{v}/v$ (where a dot indicates a derivative with respect to synchronous time $t$); they have been shown to be the suitable variables to yield an universal critical energy density in Polymer Cosmology \cite{Mantero,Patti}.

The Hamiltonian for a FLRW model with curvature filled with matter in the form of perfect fluids is
\begin{equation}
    \mathcal{H}_g(v,p_v)=-\frac{3\chi}{4\mathcal{V}}\,v\,p_v^2-\frac{3}{\chi}\,K\,c^2\,v^\frac{1}{3}\,\mathcal{V}+\rho(v)\,v\,\mathcal{V}=0,
    \label{Hamg}
\end{equation}
where $\mathcal{V}$ is an arbitrary volume constant appearing when preforming the spatial integral in the action (which however does not affect the dynamics), $\chi$ is the Einstein constant, $c$ is the speed of light, $K>0$ parameterizes the spatial curvature, and $\rho(v)=\sum_i\rho_i(v)$ contains all the necessary components, each obeying the following continuity equation that yields the following expression:
\begin{equation}
    \dot{\rho_i}+\frac{\dot{v}}{v}\,\rho_i\,(1+w_i)=0,\quad\rho_i(v)=\overline{\rho_i}\,v^{-(1+w_i)},
\end{equation}
with $w_i$ being the equation of state parameter; a simple EU model contains a radiation fluid $\rho_\gamma$ with ${w_\gamma=1/3}$ and a Cosmological Constant $\rho_\Lambda=\overline{\rho_\Lambda}$ corresponding to ${w_\Lambda=-1}$.

\begin{figure}
    \centering
    \includegraphics[width=\linewidth]{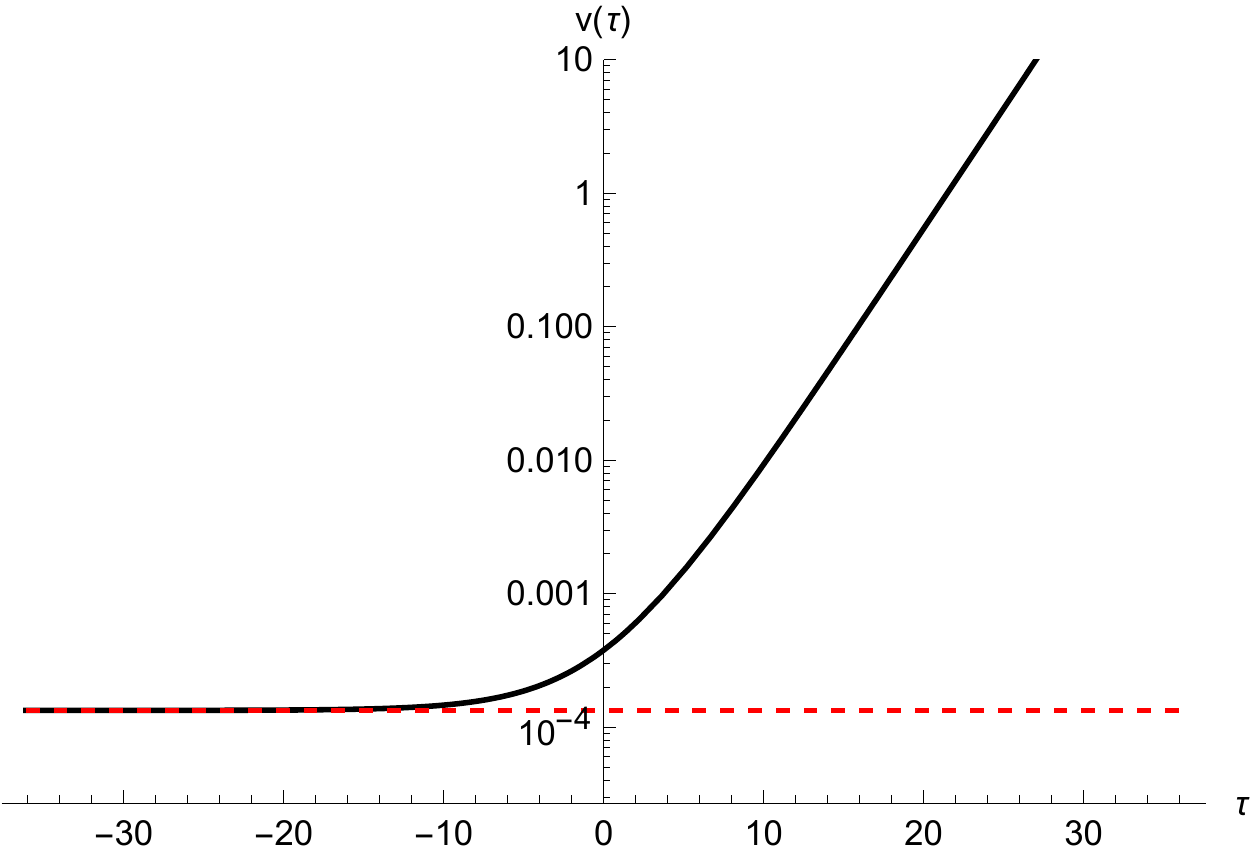}
    \caption{Evolution of the volume $v(\tau)$ in the standard EU model. The time variables are rescaled by the time $t_s$ of the beginning of standard inflation: $\tau=\frac{t}{t_s}$. The minimum value is highlighted with a red dashed line.}
    \label{standardEUgraph}
\end{figure}

From the equations of motion and the Hamiltonian constraint we obtain the Friedmann equation
\begin{equation}
    H^2=\left(\frac{\dot{v}}{3v}\right)^2=\frac{\chi}{3}(\rho_\gamma+\rho_\Lambda)-\frac{K\,c^2}{\,v^{2/3}}.
\end{equation}
By requiring the existence of a unique positive minimum $v_i$ for the volume, we obtain the following constraint on the free parameters of the densities:
\begin{equation}
    v_i=\left(\frac{3}{2}\,\frac{K\,c^2}{\chi\,\rho_\Lambda}\right)^\frac{3}{2},\quad\overline{\rho_\gamma}\,\rho_\Lambda=\left(\frac{3}{2}\,\frac{K\,c^2}{\chi}\right)^2,
    \label{finetuning}
\end{equation}
so that the Friedmann equation can be rewritten in terms of the minimum volume and easily solved:
\begin{equation}
    \dot{v}=\pm3c\,\sqrt{\frac{K}{2}\,}\,\left(\frac{v}{v_i}\right)^\frac{1}{3}\,\left(v^\frac{2}{3}-v_i^\frac{2}{3}\right),
\end{equation}
\begin{equation}
    v(t)=v_i\left[1+\exp(\pm\frac{\sqrt{2K\,}\,c\,}{\,v_i^{1/3}}\,t)\right]^\frac{3}{2}.
\end{equation}
We have two solutions, one expending to infinity and one contracting from infinity, depending on the sign of the exponential; of course we are interested in the expanding one with the $+$ sign. The solution is shown in Figure \ref{standardEUgraph}: as expected it is asymptotically Einstein static, since ${v(t\to-\infty)\to v_i>0}$, and exponentially expanding. Note that in the picture we rescaled the time variable as $\tau=t/t_s$, where $t_s\approx10^{-36}s$ is the start of inflation in the standard cosmological model; this will be useful in later sections.

Even though inflation occurs for an infinite time in the past, at any given time $t_f\gg v_i^{1/3}/\sqrt{Kc^2\,}$ there is a finite amount of $e-$folds given by
\begin{equation}
    N_e=\frac{1}{3}\,\ln(\frac{v(t_f)}{v_i})\approx\frac{\sqrt{K\,}\,c\,t_f}{\,v_i^{1/3}}.
    \label{efolds}
\end{equation}
In the next subsection we report the mechanism to realize the EU scenario.

\subsection{The Emergent Potential}
\begin{figure}
    \centering
    \includegraphics[width=\linewidth]{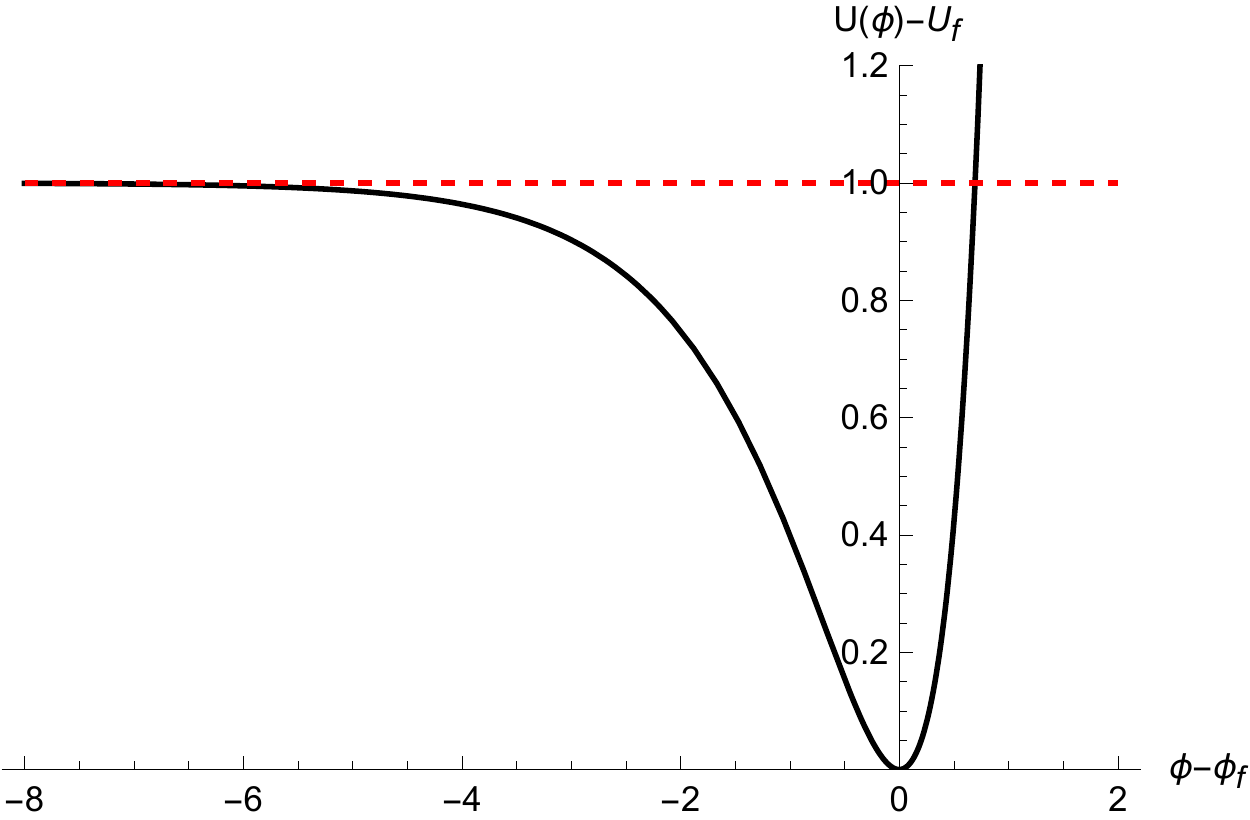}
    \caption{The potential $U(\phi)-U_f$ as function of $\phi-\phi_f$ for $U_i-U_f=1$, $\mathcal{E}=1$. The asymptote is highlighted with a red dashed line.}
    \label{potential}
\end{figure}

A simple way to create a past-infinite exponential expansion and end it at a finite time $t_f$ is to use a scalar field. Therefore we can add the scalar field term to the Hamiltonian \eqref{Hamg}:
\begin{equation}
    \mathcal{H}(v,p_v,\phi,p_\phi)=\mathcal{H}_g(v,p_v)+\rho_\phi(v,\phi,p_\phi)\,v\,\mathcal{V}=0,
\end{equation}
\begin{equation}
    \rho_\phi(v,\phi,p_\phi)=\frac{p_\phi^2}{2v^2}+U(\phi),
\end{equation}
where $U(\phi)$ is a potential and $p_\phi=\dot{\phi}\,v/c$ is the momentum conjugate to the scalar field. From the equations of motion we find that the scalar field obeys a Klein-Gordon-like equation:
\begin{equation}
    \ddot{\phi}+\frac{\dot{v}}{v}\,\dot{\phi}+c^2\pdv{U}{\phi}=0.
\end{equation}
The ideal potential for an EU model has a plateau (i.e. an asymptote) at $\phi\to-\infty$ and a well with an absolute minimum at $\phi=\phi_f$; it takes the form
\begin{equation}
    U(\phi)=U_f+(U_i-U_f)\left[\exp(\frac{\phi-\phi_f}{\mathcal{E}})-1\right]^2,
\end{equation}
where $U_i$ is the asymptotic value in the infinite past, $U_f$ is the minimum value and $\mathcal{E}$ is a constant scale parametrizing the width of the well. The form of the potential is shown in Figure \ref{potential} for generic values of the parameters.

For $t\to-\infty$ we have $\phi\to-\infty$ and the field is in the plateau of the potential; this implies $\dot{\phi}^2\ll U$ and therefore the energy density $\rho_\phi\approx U_i$ is nearly constant, playing the role of the Cosmological Constant $\rho_\Lambda$ of the previous subsection. When we approach $t=t_f$ the field falls into the well until it reaches the minimum $U_f\ll U_i$, and the exponential expansion ends. (Note that it is possible to set $U_f\neq0$ to represent the late-time cosmological constant \cite{OriginalEU}, but this is beyond the scope of this study.)

As mentioned before, the infinite time of inflation produces a finite amount of expansion; provided that $v_i$ and $t_f$ are respectively chosen small and large enough, a very large amount of $e-$folds can be produced as follows from equation \eqref{efolds}, thus solving all the paradoxes of Friedmann evolution in a similar manner to the standard inflationary theory \cite{MontaniPrimordial}. However, analogously to the latter, this model is also subject to some form of fine-tuning and criticisms.

\subsection{Fine-Tuning}
As mentioned above, some fine-tuning is needed in the EU to reproduce observational parameters, such as density perturbations of the order $\order{10^{-5}}$ and a late-time Cosmological Constant $\Omega_\Lambda\approx0.7$; however, all inflationary universe models need some amount of fine-tuning. The specific geometrical fine-tuning problem in the EU models is the requirement of a particular choice of the initial volume $v_i$ and of the primordial cosmological constant $\rho_\Lambda$ or $U_i$. This choice must then be supplemented by a further fine-tuning: a choice of initial kinetic energy such that the inequality $\dot{\phi}^2\ll U_i$ holds. Both conditions are required to attain an asymptotically Einstein-static state.

The authors of \cite{OriginalEU,EU2} acknowledge the necessity of fine-tuning in this model, but claim that the situation is not too different from any other inflationary model. Besides, they argue that the advantages of having a non-singular, highly symmetric initial state overcome the troubles of fine-tuning. However, the scope of this work is to provide a mechanism to generate an EU model with the minimum fine-tuning necessary.

\section{Emergent Universe from a Modified Algebra}
\label{EUpoly}
In this section we present a modified Heisenberg algebra, that in the classical limit translates to modified Poisson brackets, which is able to yield an EU-like solution without the need of much fine-tuning.

The modified algebra, inspired by quantum gravity and quantum cosmological theories such as PQM \cite{CorichiPQM,Battisti} and the GUP representation \cite{Maggiore,Kempf,BarcaGUP,SebastianoGUP,BossioGUP,FadelMaggiore}, takes the form
\begin{equation}
    \comm{\hat{q}}{\hat{p}}=i\hslash\,\left(1-\frac{\mu^2\ell_P^2\hat{p}^2}{\hslash^2}\right),
    \label{commutator}
\end{equation}
where $\hat{q}$ and $\hat{p}$ are operators corresponding to two generic conjugate variables and $\mu>0$ is a free real parameter that is reminiscent of the lattice spacing in PQM but here takes the role of just a deformation parameter similarly to the GUP representation. When implementing these modified commutation relations (that in the semiclassical setting will become modified Poisson brackets), we will insert appropriate fundamental constant in order to always have $\mu$ as a dimensionless parameter, as is sometimes done in GUP literature \cite{GUPbeta0,FadelMaggiore}; for example, in the commutator \eqref{commutator} we assumed $q$ and $p$ to be the standard position and momentum respectively, so we inserted the Planck length $\ell_P$ and the reduced Planck constant $\hslash$ to keep both the term in parentheses and the deformation parameter $\mu$ dimensionless.

We will see how this algebra, when implemented on the cosmological minisuperspace at a (semi)classical level, will lead to an avoidance of the Big Bang singularity (similarly to Polymer Cosmology \cite{Mantero,Review}) with the introduction of an asymptotic minimum, as already mentioned in \cite{BarcaGUP}.

\subsection{A Simple Example}
In the classical limit, the commutator \eqref{commutator} becomes a rule for Poisson brackets. In this first example, we will not consider curvature and will not assume any specific kind of matter but leave a generic energy density ${\rho(v)=\overline{\rho\,}\,v^{-(1+w)}}$.

\begin{figure}
    \centering
    \includegraphics[width=\linewidth]{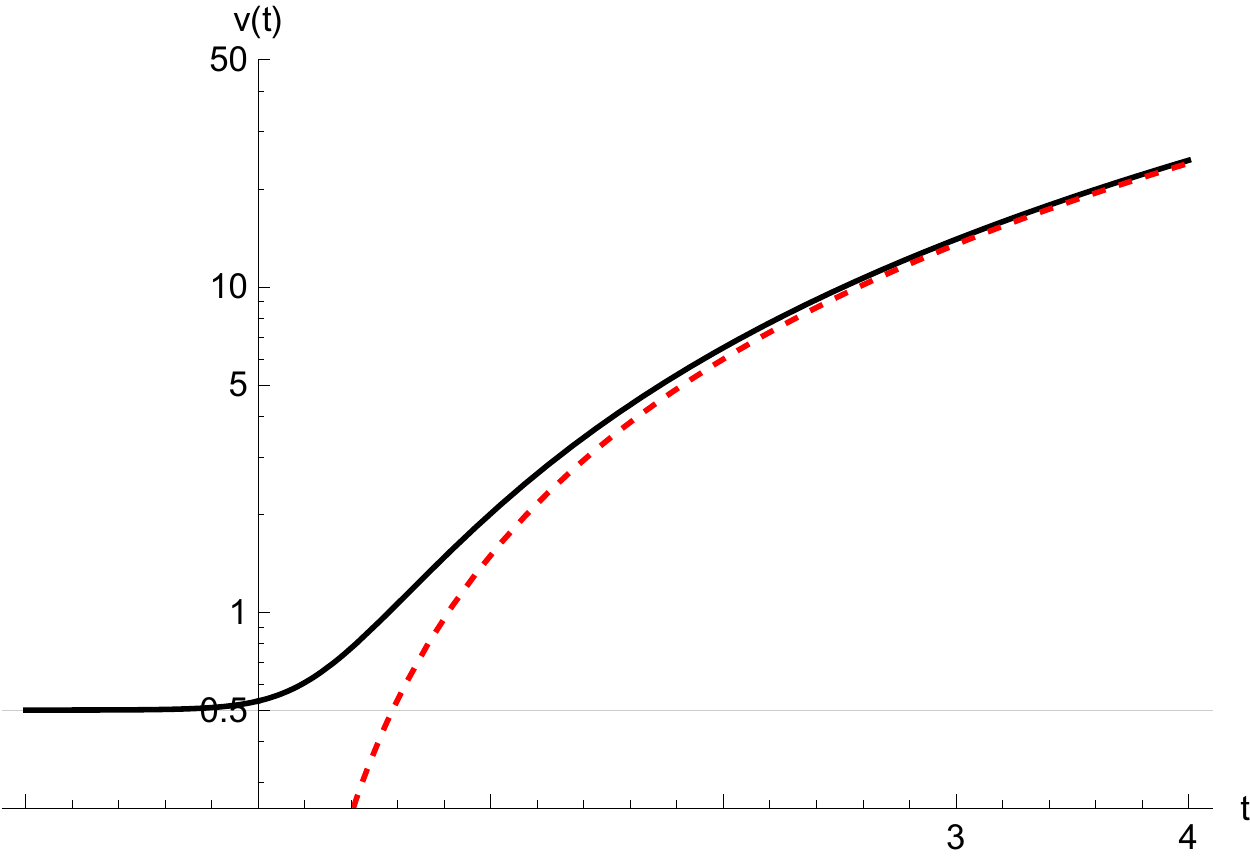}
    \caption{The asymptotic solution $v(t)$ for the simple model with a generic energy density (black continuous line), compared with the standard evolution (red dashed line) which falls into the singularity. The asymptotic volume $v_i$ is highlighted with the grey faded line.}
    \label{atgh}
\end{figure}

The Hamiltonian constraint is the same as \eqref{Hamg}, but with no curvature and modified Poisson brackets:
\begin{equation}
    \mathcal{H}_g(v,p_v)=-\frac{3\chi}{4\mathcal{V}}\,v\,p_v^2+\rho(v)\,v\,\mathcal{V}=0,
\end{equation}
\begin{equation}
    \pb{v}{p_v}=1-\frac{\mu^2p_v^2}{\hslash^2};
\end{equation}
note that, since the volume $v$ is dimensionless, $p_v$ has the dimensions of an action and therefore we divided the correction term by $\hslash^2$ to keep $\mu$ dimensionless. Then from the equations of motion and the constraint we derive a modified Friedmann equation:
\begin{equation}
    H^2=\frac{\chi}{3}\rho\left(1-\frac{\rho}{\,\rho_\mu}\right)^2,\quad\rho_\mu=\frac{3\chi\hslash^2}{4\mu^2\mathcal{V}^2},
    \label{modFried}
\end{equation}
where $\rho_\mu$ is a critical energy density that is constant \cite{Mantero} and introduces a critical point on the dynamics; the critical point is calculated as the value $v_i$ such that $\dot{v}=0$, which, as long as ${w\neq-1}$, implies
\begin{equation}
    1-\frac{\rho(v_i)}{\,\rho_\mu}=0,\quad v_i=\left(\frac{\overline{\rho\,}}{\rho_\mu}\right)^{\frac{1}{1+w}}.
    \label{vminPUP}
\end{equation}
The solution $v(t)$ then has the following implicit form:
\begin{equation}
    \left(\frac{v(t)}{v_i}\right)^\frac{1+w}{2}-\,\text{atanh}\left(\Big(\frac{v(t)}{v_i}\Big)^{-\frac{1+w}{2}}\right)=\pm\frac{1+w}{2}\,t\,\sqrt{3\,\rho_\mu\,\chi\,}\,;
    \label{v(t)PUP}
\end{equation}
again we have two solutions, one contracting and one expanding, depending on the sign. The solution of interest (the expanding one with the $+$ sign) is shown in Figure \ref{atgh} for generic values of the parameters. Of course this does not present an exponential behaviour, since at this stage we did not include a Cosmological Constant; however this is just a simplified model to show the ability of the modified algebra \eqref{commutator} to naturally implement an asymptotic minimum value.

The main result of this simple construction is that we did not have to impose any fine-tuning such as the constraint \eqref{finetuning} in order to obtain a positive minimum for the volume; it naturally follows from the form of the correction factor $(1-\frac{\rho}{\rho_\mu})^2$ in the modified Friedmann equation \eqref{modFried}. We obtain a non-singular, asymptotically Einstein-static model that in the future yields the standard Friedmann evolution; indeed, note that for $v\gg v_i$ we have $\rho(v)\ll\rho_\mu$, and the modified Friedmann equation reduces to the standard one $H^2=\chi\,\rho/3$; this can be also seen from equation \eqref{v(t)PUP}: in the limit $v\gg v_i$ corresponding to $t\to+\infty$, the argument of the hyperbolic arctangent goes to zero and, given the relation \eqref{vminPUP} between $\rho_\mu$, $\overline{\rho}$ and $v_i$, we obtain the standard Friedmann evolution
\begin{equation}
    v(t)=\left(\sqrt{3\,\overline{\rho}\,\chi\,}\,\frac{1+w}{2}\,t\right)^\frac{2}{1+w}.
\end{equation}

In the following subsection we will implement this scheme on the full model with curvature and a Cosmological Constant coming from a slow-rolling phase of a scalar field as in the previous Section \ref{ham}.

\subsection{The Full Model}
We will now consider the full model. We will consider different phases: the first, near the classical singularity, where the matter-energy is dominated by a relativistic component; the second where a scalar field potential grows, yielding an inflationary phase dominated by a Cosmological Constant; a final one where the scalar field has again decayed into photons and the late-time evolution becomes Friedmann-like. In all phases we will consider positive curvature, even though in the modified algebra scheme it is not needed to obtain an asymptotic behaviour, in order to make the comparison with the standard EU model more immediate.

The full Hamiltonian of the model is
\begin{equation}
    \mathcal{H}(v,p_v,\phi,p_\phi)=-\frac{3\chi}{4\mathcal{V}}\,\chi\,v\,p_v^2-\frac{3}{\chi}\,K\,c^2\,v^\frac{1}{3}\,\mathcal{V}+\rho\,v\,\mathcal{V}=0,
\end{equation}
\begin{align}
    &\text{phase }1)\quad\rho=\rho_\gamma=\overline{\,\rho_\gamma^\text{pre}\,\,}\,v^{-\frac{4}{3}},\\
    &\text{phase }2)\quad\rho=\rho_\phi(U\gg\dot{\phi}^2)=U_i=\rho_\Lambda,\\
    &\text{phase }3)\quad\rho=\rho_\gamma=\overline{\,\rho_\gamma^\text{post}\,\,}\,v^{-\frac{4}{3}},
\end{align}
where the constants $\overline{\rho_\gamma^\text{pre/post}\,}$ and $\rho_\Lambda$ have been chosen to maintain continuity for $v$ and $\dot{v}$.

\begin{figure}
    \centering
    \includegraphics[width=\linewidth]{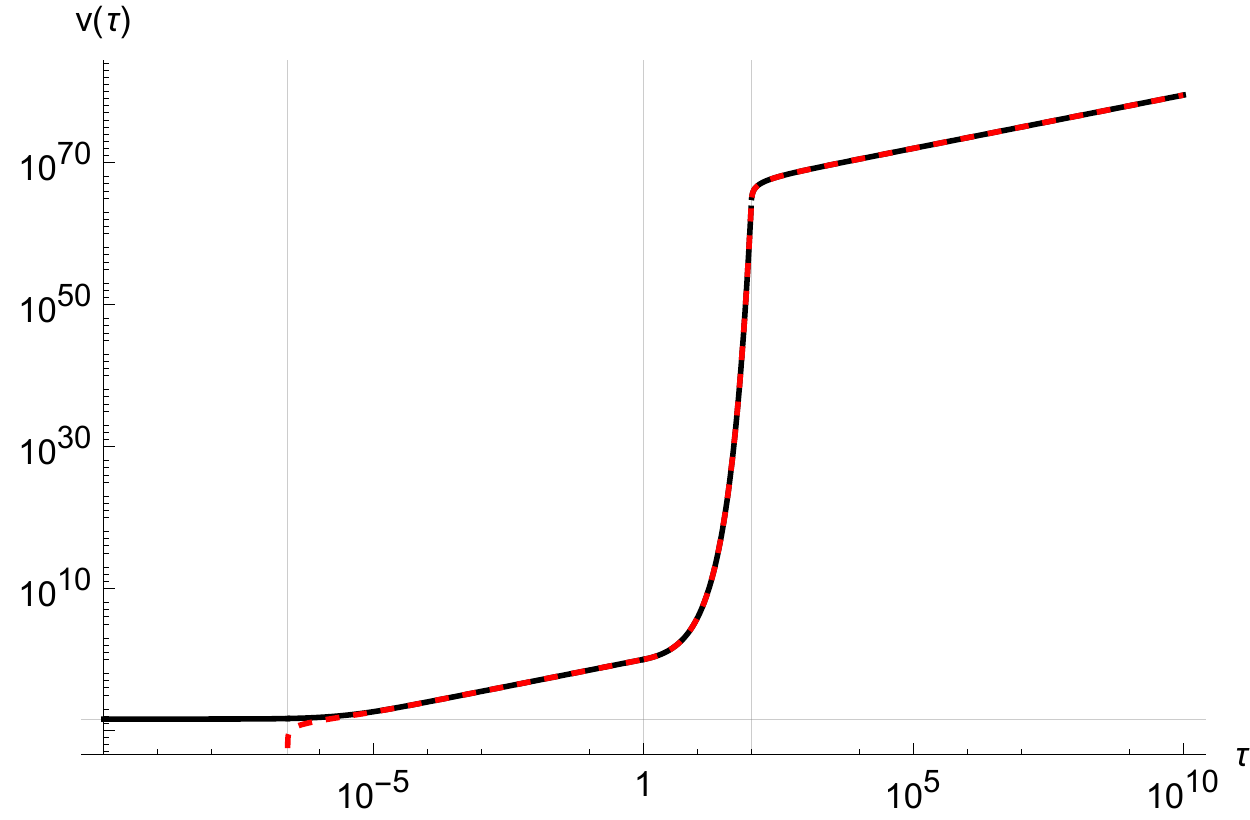}
    \caption{The evolution of $v(\tau)$ for the full model (black continuous line) compared with the standard dynamics (dashed red line). The minimum volume $v_i$ is highlighted with a grey faded horizontal line, while the grey faded vertical lines separate the different phases (from left to right they indicate the classical Big Bang, the start of inflation and its end).}
    \label{fullmodel}
\end{figure}

\begin{figure}
    \centering
    \includegraphics[width=\linewidth]{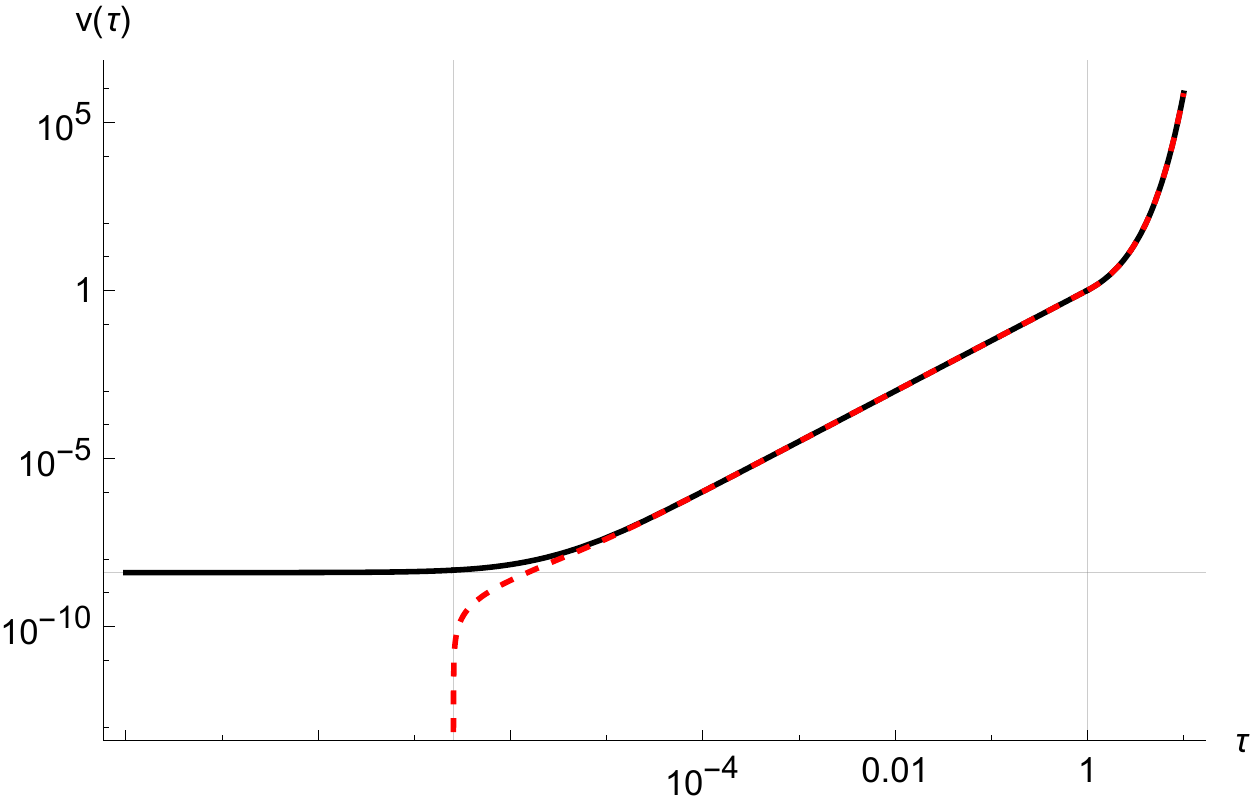}
    \caption{Zoomed-in version of Figure \ref{fullmodel} to give a better view of the behaviour near the classical singularity.}
    \label{fullmodelzoom}
\end{figure}

Given the complexity of the corresponding Friedmann equations, the resolution has been performed numerically. Again, we rescaled all quantities by their corresponding value at the beginning of inflation, that is, we used as time variable $\tau=t/t_s$ and all densities have been rescaled accordingly. The result is shown in Figure \ref{fullmodel} for the whole evolution and compared with the classical case (i.e. the one obtained with standard Poisson brackets $\pb{v}{p_v}=1$); Figure \ref{fullmodelzoom} is the same picture zoomed near the singularity, to better highlight the asymptotic behaviour.


We see that we did not have to impose any condition such as \eqref{finetuning} in order to obtain an asymptotic behaviour, it is implemented naturally by the modified algebra \eqref{commutator}. Besides, the standard dynamics is recovered pretty soon and already shortly before the inflationary epoch the evolution is practically indistinguishable; this will allow us to use the classical Friedmann equation for Inflation when in the next section we will calculate the primordial Power Spectrum.

\section{Modified Power Spectrum of Perturbations}
\label{power}
In this section, as a phenomenological consequence of this model and in particular of the modified algebra \eqref{commutator}, we aim to derive the modified Power Spectrum of primordial scalar perturbations during the inflationary epoch. We will partially follow \cite{Kiefer} but compute the spectrum through a different method. For other approaches for the computation of quantum gravity corrections to the inflationary spectrum, see \cite{Maniccia,Torcellini}.

The general action for a scalar field takes the form
\begin{equation}
    S_\phi=\int\frac{dt\,d^3x}{2}\sqrt{-g\,}\,\Bigl(g^{\mu\nu}\partial_\mu\phi\partial_\nu\phi-2U(\phi)\Bigr);
    \label{phitotalaction}
\end{equation}
introducing conformal time $\eta$, we can rewrite the zero-order homogeneous action for the scalar field:
\begin{equation}
    d\eta=\frac{dt}{a},
\end{equation}
\begin{equation}
    \hspace{-0.7cm}S_\phi=\int\frac{dt}{2}a^3\mathcal{L}^3\left(\frac{\dot{\phi}^2}{c^2}-2U(\phi)\right)=\int\frac{d\eta}{2}\mathcal{L}^3\Bigl(a^2\frac{(\phi')^2}{c^2}-2a^4U(\phi)\Bigr),
\end{equation}
where a prime denotes a derivative with respect to $\eta$ and $\mathcal{L}$ is an arbitrary length scale that appears when performing the volume integral.

At this point it is useful to introduce the so-called Mukhanov-Sasaki variable $\xi$, a master gauge-invariant variable which is sufficient to fully describe the scalar sector of perturbations \cite{Mukhanov}:
\begin{equation}
    \xi(x,\eta)=a\left(\delta\phi_\text{GI}+\frac{\phi'\,\Phi_\text{B}}{aH}\right),
\end{equation}
where $\delta\phi_\text{GI}$ is the gauge-invariant form of the scalar field perturbations and $\Phi_\text{B}$ is a Bardeen potential depending on the perturbative scalar functions in the perturbed metric \cite{Kiefer}. The action for the variable $\xi$ is obtained as the scalar part of the second variation of the total action (that is, of both the gravitational sector and the matter action \eqref{phitotalaction}) \cite{Mukhanov}:
\begin{equation}
    \delta^2S=\int\frac{d\eta\,d^3x}{c^2}\left[(\xi')^2-\delta^{ij}\,c^2\,\partial_i\xi\,\partial_j\xi+\xi^2\,\frac{z''}{z}\right],
\end{equation}
\begin{equation}
    z=a\sqrt{\epsilon\,},\quad\epsilon=-\frac{\dot{H}}{H^2},
\end{equation}
where $\epsilon$ is the first slow-roll parameter.

Now, since we are working with linear perturbations where each mode evolves independently, we can perform a Fourier decomposition so that the action greatly simplifies:
\begin{equation}
    \xi(x,\eta)=\frac{1}{\mathcal{L}^\frac{3}{2}}\,\sum_k\xi_k(\eta)e^{ikx},
\end{equation}
\begin{equation}
    \delta^2S=\int\frac{d\eta}{c^2}\,\sum_k\Bigl(\xi^*_k\,'\,\xi_k'-\omega_k^2(\eta)\xi_k^*\,\xi_k\Bigr),
\end{equation}
where we have defined a frequency
\begin{equation}
    \omega_k^2(\eta)=k^2c^2-\frac{z''}{z}.
\end{equation}
Note that if we assume $\xi$ to be real, then we will have ${\xi_k^*=\xi_{-k}}$. The momentum conjugate to $\xi_k$ is defined as ${\pi_k=\xi_k'/c^2}$ and we finally obtain the Hamiltonian for the scalar perturbations:
\begin{equation}
    \mathcal{H}=\sum_k\mathcal{H}_k=\sum_k\frac{c^2}{2}\pi_k^*\pi_k+\frac{\omega_k^2(\eta)}{2c^2}\xi_k^*\xi_k.
\end{equation}

In order to calculate the Power Spectrum, we will make the assumption that during the inflationary era the evolution is dominated by the Cosmological Constant and therefore all other components are negligible; besides, if inflation starts late enough, we will have $\rho\ll\rho_\mu$ and, as mentioned in the last section, we can neglect the correction factor in \eqref{modFried}:
\begin{equation}
    H^2=\frac{\chi}{3}\,\rho_\Lambda=H_s^2,
\end{equation}
where $H_s$ is the constant Hubble parameter of inflation. Note that, to be precise, in a pure de Sitter universe the background matter field is set to a constant value and thus, in principle, it does not make sense to speak about its perturbations. This is shown explicitly in the appearance of the slow-roll parameter $\epsilon$, which in this limit should be vanishing; indeed in this case a Power Spectrum cannot be obtained since inflation never stops. Nonetheless, the computations can be performed by keeping the slow-roll parameter as a non-vanishing constant, and this particular case represents a very good and easy-to-compute
example to derive a Power Spectrum. In this regime the conformal time acquires a precise dependence on the scale factor and the frequency $\omega_k$ greatly simplifies:
\begin{equation}
    \eta=-\frac{1}{aH_s},
    \label{eta}
\end{equation}
\begin{equation}
    \omega_k^2(\eta)=k^2c^2-\frac{2}{\eta^2}.
\end{equation}

Before implementing quantization, we should in principle define real analogues of $\xi_k$ and $\pi_k$, otherwise the procedure is not entirely consistent \cite{QuantumMeasurement}. However this will make no difference in later calculations, and therefore we will not define such new variables to avoid cluttering the notation, as done in \cite{Kiefer}. The only modification that we will implement is a rescaling of both variables by the speed of light $c$ to simplify the constraint; in particular we will substitute ${\pi_k^\text{new}=\pi_k^\text{old}c}$ and ${\xi_k^\text{new}=\xi_k^\text{old}/c}$. Therefore the Hamiltonian operator that we will use is
\begin{equation}
    \hat{\mathcal{H}}=\sum_k\hat{\mathcal{H}}_k=\sum_k\frac{\hat{\pi}_k^2}{2}+\frac{\omega_k^2(\eta)}{2}\hat{\xi}_k^2;
\end{equation}
we will recover the right units later in the definition of the Power Spectrum.

Now we can perform the quantization of the system and proceed to compute the Power Spectrum. We will first briefly present the standard Spectrum derived through the canonical quantization, and then find the modified spectrum coming from the algebra \eqref{commutator}.

\subsection{Standard Power Spectrum}
Here we will compute the standard Power Spectrum. In the standard representation of Quantum Mechanics, the two operators corresponding to the Fourier modes will obey the standard commutation relations and will have the standard action:
\begin{equation}
    \comm{\hat{\xi}_k}{\hat{\pi}_k}=i\hslash,
\end{equation}
\begin{equation}
    \hat{\xi}_k\,\psi(\xi_k)=\xi_k\,\psi(\xi_k),\quad\hat{\pi}_k\,\psi(\xi_k)=-i\hslash\pdv{\xi_k}\psi(\xi_k).
\end{equation}

A single Fourier mode has Hamiltonian $\mathcal{H}_k$ with a time-dependent frequency $\omega_k(\eta)$; therefore the wavefunctions $\psi(\eta,\xi_k)$ will obey a time-dependent Schr\"odinger equation of the form
\begin{equation}
    i\hslash\,\pdv{\eta}\psi(\eta,\xi_k)=\frac{1}{2}\left(-\hslash^2\pdv[2]{\xi_k}+\omega_k^2(\eta)\xi_k^2\right)\psi(\eta,\xi_k).
    \label{timedependentSchroedingerxi}
\end{equation}
This is the Schr\"odinger equation of a harmonic oscillator with time-dependent frequency. The solution to such a system can be found through the method of invariants \cite{Invariants1,Invariants2,Invariants3} and is a superposition of the following normalized wavefunctions:
\begin{equation}
    \psi_n(\eta,\xi_k)=\frac{h_n(\frac{\xi_k}{\sqrt{\hslash\,}\,f})}{\sqrt{2^n\,n!\,}}\,\frac{e^{-\frac{\xi_k^2}{2\hslash f^2}}}{(\pi\hslash f^2)^\frac{1}{4}}\,e^{i\,\frac{\,\,f'}{2\hslash f}\,\xi_k^2}\,e^{i\alpha_n},
    \label{psin}
\end{equation}
where $\alpha_n=\alpha_n(\eta)=-(n+\frac{1}{2})\int f^{-2}d\eta$ is a time-dependent phase, $h_n$ are Hermite polynomials and ${f=f(\eta)}$ is an auxiliary function with the dimension of the square root of time that is the solution of the following differential equation:
\begin{equation}
    f''+\omega_k^2f-f^{-3}=0;
    \label{auxiliary}
\end{equation}
the solution to the time-independent harmonic oscillator can be easily recovered by making the substitution ${f\to1/\sqrt{\omega_k}}$ and making it constant.

The Spectrum for $\xi_k$ can then be calculated by linking its perturbations to the curvature perturbations \cite{Kiefer}, yielding
\begin{equation}
    \mathcal{P}^\text{std}(k)=\eval{\frac{c^2k^3}{4\pi^2}\,\frac{\ev{\hat{\xi}_k^2}{0}}{a^2\epsilon}}_{-ck\eta\ll1}
\end{equation}
where $\eta\to0^-$ corresponds to $t\to+\infty$ so that $-ck\eta\to0$ is the large scale limit (the factor $c^2$ appears due to the rescaling of $\xi_k$ performed earlier). The expectation value of $\hat{\xi}_k^2$ is computed on the vacuum state i.e. the ground state of the time-dependent oscillator; we therefore need to know how to express the result of $\hat{\xi}_k\psi_n$. This can be done by constructing ladder operators for the time-dependent system: they take the form \cite{TDHOcoherent}
\begin{equation}
    \hat{a}^\dagger=\frac{\frac{\hat{\xi}_k}{f}-i(f\hat{\pi}_k-f'\hat{\xi}_k)}{\sqrt{2\,\hslash\,}},\quad\hat{a}^\dagger\psi_n=\sqrt{n+1\,}\,e^{i\varphi}\psi_{n+1};
\end{equation}
\begin{equation}
    \hat{a}=\frac{\frac{\hat{\xi}_k}{f}+i(f\hat{\pi}_k-f'\hat{\xi}_k)}{\sqrt{2\,\hslash\,}},\quad\hat{a}\psi_n=\sqrt{n\,}\,e^{-i\varphi}\psi_{n-1};
\end{equation}
from these we can derive the expressions of $\hat{\xi}_k$ and $\hat{\pi}_k$ as functions of the ladder operators, and their actions on an eigenstate $\psi_n$:
\begin{equation}
    \hat{\xi}_k=\sqrt{\hslash\,}\,f\,\frac{\hat{a}^\dagger+\hat{a}}{\sqrt{2}},
\end{equation}
\begin{equation}
    \hat{\pi}_k=i\,\frac{\sqrt{\hslash\,}}{f}\,\frac{\hat{a}^\dagger-\hat{a}}{\sqrt{2}}+\sqrt{\hslash\,}\,f'\,\frac{\hat{a}^\dagger+\hat{a}}{\sqrt{2}},
\end{equation}
\begin{equation}
    \hat{\xi}_k\psi_n=\sqrt{\hslash\,}\,f\left(\sqrt{\frac{n+1}{2}\,}\,e^{i\varphi}\psi_{n+1}+\sqrt{\frac{n}{2}\,}\,e^{-i\varphi}\psi_{n-1}\right),
    \label{resultofxik}
\end{equation}
\begin{equation}
    \hat{\pi}_k\psi_n=i\frac{\sqrt{\hslash\,}}{f}\left(R\sqrt{\frac{n+1}{2}\,}\,e^{i\varphi}\psi_{n+1}-R^*\sqrt{\frac{n}{2}\,}\,e^{-i\varphi}\psi_{n-1}\right),
    \label{resultofpik}
\end{equation}
where we defined
\begin{equation}
    R=1-iff',\quad\varphi=\int\frac{d\eta}{f^2(\eta)}.
    \label{defineR}
\end{equation}
Finally, we can write the single-mode Hamiltonian operator as function of the ladder operators, and find its action on a state $\psi_n$:
\begin{equation}
\begin{aligned}
    \hat{\mathcal{H}}_k=&\frac{\hslash}{4f^2}\Bigl(\omega_k^2f^4(\hat{a}^\dagger+\hat{a})^2-(\hat{a}^\dagger-\hat{a})^2\Bigr)+\\
    +&\frac{\hslash}{4}\,\frac{f'}{f}\,(\hat{a}^\dagger+\hat{a})\Bigl(f\,f'(\hat{a}^\dagger+\hat{a})+2i(\hat{a}^\dagger-\hat{a})\Bigr),
\end{aligned}
\end{equation}
\begin{equation}
\begin{aligned}
    \hat{\mathcal{H}}_k\psi_n=&\frac{\hslash}{4f^2}\,(2n+1)\Bigl(f^4\omega_k^2+f^2f'^2+1\Bigr)\psi_n+\\
    +&\frac{\hslash}{4f^2}\,\sqrt{(n+1)(n+2)\,}\,e^{2i\varphi}\Bigl(f^4\omega_k^2-R^2\Bigr)\psi_{n+2}+\\
    +&\frac{\hslash}{4f^2}\,\sqrt{n(n-1)\,}\,e^{-2i\varphi}\Bigl(f^4\omega_k^2-(R^*)^2\Bigr)\psi_{n-2}.
\end{aligned}
\end{equation}
Obviously this expression implies that the states $\psi_n$ are not eigenstates of the Hamiltonian operator, which was to be expected since it is explicitly time-dependent; however, if we again make the substitution $f\to1/\sqrt{\omega_k\,}$ and make it a constant, so that ${f'=0}$, ${R=R^*=1}$, and ${f^4\omega_k^2=1}$, all these relations reduce to the standard formulas of the time-independent harmonic oscillator, including ${\hat{\mathcal{H}}_k=\hslash\omega_k\hat{a}^\dagger\hat{a}}$. Nevertheless, in the time-dependent system it is still possible to construct the operator
\begin{equation}
    \hat{I}=\hslash\,\hat{a}^\dagger\,\hat{a}=\frac{\frac{\hat{\xi_k}^2}{f^2}+(f\hat{\pi}_k-f'\hat{\xi}_k)^2}{2},\quad\hat{I}\psi_n=\hslash\bigl(n+\frac{1}{2}\bigr)\psi_n,
\end{equation}
which is actually the quantum version of the original invariant defined by Lewis and Riesenfeld \cite{Invariants1,Invariants2}, and it can be used to find coherent states for the time-dependent harmonic oscillator; they reduce to the standard coherent states of the time-independent harmonic oscillator under the substitution ${f=1/\sqrt{\omega_k\,}=\,}$const. \cite{TDHOcoherent}. As a final comment, we note that all these relations in term of the ladder operators are the same regardless if the states $\psi_n$ are expressed in the $\xi_k$ or the $\pi_k$ polarization.

Now we can proceed to compute the expectation value of $\xi_k^2$ on the ground state. Since ${\hat{\xi}_k\psi_0=e^{i\varphi}\sqrt{\hslash\,}\,f\,\psi_1/\sqrt{2\,}}$, we have
\begin{equation}
\begin{aligned}
    \ev{\hat{\xi}_k^2}{0}&=\int_{-\infty}^{+\infty}d\xi_k\,\psi_0^*\hat{\xi}_k^2\psi_0=\int_{-\infty}^{+\infty}d\xi_k\abs{\hat{\xi}_k\psi_0}^2=\\
    &=\int_{-\infty}^{+\infty}d\xi_k\abs{\frac{\sqrt{\hslash\,}\,f\,\psi_1\,e^{i\varphi}}{\sqrt{2\,}}}^2=\frac{\hslash}{2}\,f^2(\eta).
    \label{csiquadrostandard}
\end{aligned}
\end{equation}
To calculate the spectrum we just need to find the expression of $f(\eta)$.

The solution to the auxiliary equation \eqref{auxiliary} can be constructed from the solutions $f_1$ and $f_2$ of the corresponding homogeneous equation:
\begin{equation}
    f''+\omega_k^2f=0,
\end{equation}
\begin{equation}
    f_1(\eta)=\frac{1}{\sqrt{ck\,}\,}\left(\cos(ck\eta)-\frac{\sin(ck\eta)}{ck\eta}\right),
\end{equation}
\begin{equation}
    f_2(\eta)=\frac{1}{\sqrt{ck\,}\,}\left(\frac{\cos(ck\eta)}{ck\eta}+\sin(ck\eta)\right).
\end{equation}
Then the function $f$ takes the form
\begin{equation}
    f(\eta)=\frac{1}{\mathcal{W}}\,\biggl(A_1^2f_1^2+A_2^2f_2^2+2f_1f_2\sqrt{A_1^2A_2^2-\mathcal{W}^2\,}\biggr)^\frac{1}{2},
\end{equation}
where $A_1$, $A_2$ are $\eta$-independent constants and $\mathcal{W}$ is the Wronskian:
\begin{equation}
    \mathcal{W}=f_1f_2'-f_1'f_2=1.
\end{equation}
The two constants must be set through initial conditions: we require that at the beginning of inflation, when all the modes of astrophysical interest today have a physical wavelength smaller than the Hubble radius $\frac{ck}{aH}\gg1$, the expansion of the Universe does not affect perturbations and therefore each mode behaves as a harmonic oscillator with constant frequency. Hence we impose that modes asymptotically approach Minkowskian quantum harmonic oscillators with frequency $ck$:
\begin{equation}
    \lim_{-ck\eta\to\infty}f(\eta)=\frac{1}{\sqrt{ck\,}}\,;
\end{equation}
this is satisfied by setting $A_1^2=A_2^2=1$, so that the expression for $f$ is
\begin{equation}
    f(\eta)=\sqrt{\frac{1+c^2k^2\eta^2}{c^3k^3\eta^2}\,}.
    \label{rho}
\end{equation}

Then, inserting this expression into the Spectrum, taking the large scale limit $-ck\eta\ll1$ and remembering the dependence of $\eta$ on the scale factor \eqref{eta}, the final expression for the spectrum is
\begin{equation}
\begin{aligned}
    \mathcal{P}^\text{std}(k)=\eval{\frac{c^2k^3}{4\pi^2}\,\frac{\hslash f^2(\eta)}{2a^2\epsilon}}_{-ck\eta\ll1}=&\\
    =\eval{\frac{\hslash}{c}\,\frac{H_s^2}{8\pi^2\epsilon}(1+c^2k^2\eta^2)}_{-ck\eta\ll1}=&\,\frac{\hslash}{c}\,\frac{H_s^2}{8\pi^2\epsilon}.
\end{aligned}
\end{equation}
We have obtained the usual flat, $k$-independent Spectrum \cite{Weinberg}.

\subsection{Modified Power Spectrum}
Here we will derive the Power Spectrum that arises from the Fourier-transformed Mukhanov-Sasaki variable $\xi_k$ obeying the modified algebra \eqref{commutator}:
\begin{equation}
    \comm{\hat{\xi}_k}{\hat{\pi}_k}=i\hslash\,(1-\frac{\mu^2t_P\hat{\pi}_k^2}{\hslash}),
\end{equation}
where $\pi_k^2$ has the dimensions of an energy so we introduced the Planck constant and Planck time $t_P$ to keep the deformation parameter $\mu$ still dimensionless. Due to the modified commutator depending on $\pi_k$, it will be easier to work in the momentum polarization, i.e. with wavefunctions $\psi=\psi(\eta,\pi_k)$.

By using arguments similar to those in \cite{BarcaGUP,SebastianoGUP}, if we impose that in the momentum polarization the scalar field operator acts simply differentially, we can find the action of the multiplicative momentum operator ${\hat{\pi}_k\psi(\pi_k)=g(\pi_k)\psi(\pi_k)}$ as
\begin{equation}
    \dv{g}{\pi_k}=1-\frac{t_P}{\hslash}\,\mu^2\,g^2,\quad\sqrt{\frac{\hslash}{t_P}\,}\,\frac{\text{atanh}(\sqrt{\frac{t_P}{\hslash}\,}\,\mu\,g)}{\mu}=\pi_k;
    \label{modactions}
\end{equation}
therefore the action of the fundamental operators is
\begin{align}
    \hat{\pi}_k\,\psi=&\sqrt{\frac{\hslash}{t_P}\,}\,\frac{\tanh(\sqrt{\frac{t_P}{\hslash}\,}\,\mu\,\pi_k)}{\mu}\,\psi,\label{actionp}\\
    \hat{\xi}_k\,\psi=&i\hslash\,\pdv{\pi_k}\psi.
\end{align}

Given the action \eqref{actionp} for the modified operator $\hat{\pi}_k$, the Hamiltonian $\mathcal{H}_k$ for a single Fourier mode yields a time-dependent Schr\"odinger equation with a modified kinetic term:
\begin{equation}
    i\hslash\,\pdv{\eta}\psi=\frac{1}{2}\left(\frac{\hslash}{t_P}\,\frac{\tanh[2](\sqrt{\frac{t_P}{\hslash}\,}\,\mu\,\pi_k)}{\mu^2}-\hslash^2\omega_k^2(\eta)\,\pdv[2]{\pi_k}\right)\psi.
    \label{schroedingertanh}
\end{equation}
This partial differential equation (PDE) is quite difficult to solve, so we perform an expansion in powers of $\mu^2$:
\begin{equation}
    \frac{\hslash}{t_P}\,\frac{\tanh[2](\sqrt{\frac{t_P}{\hslash}\,}\,\mu\,\pi_k)}{\mu^2}=\pi_k^2-\mu^2\,\frac{t_P}{\hslash}\,\frac{2\pi_k^4}{3}+\order{\mu^4},
\end{equation}
\begin{equation}
    \psi(\eta,\pi_k)=\psi^0(\eta,\pi_k)+\mu^2\,\psi^1(\eta,\pi_k)+\order{\mu^4}.
\end{equation}
Plugging these expansions back into the Schr\"odinger equation \eqref{schroedingertanh} and separating the different powers of $\mu^2$, we obtain two new PDEs for the two components $\psi^0$ and $\psi^1$:
\begin{align}
    &i\hslash\,\pdv{\eta}\psi^0=\frac{1}{2}\left(\pi_k^2-\hslash^2\omega_k^2(\eta)\,\pdv[2]{\pi_k}\right)\psi^0,\label{TDSEzeroorder}\\
    &i\hslash\,\pdv{\eta}\psi^1=\frac{1}{2}\left(\pi_k^2-\hslash^2\omega_k^2(\eta)\,\pdv[2]{\pi_k}\right)\psi^1+F\label{TDSEfirstorder},
\end{align}
\begin{equation}
    F=F(\eta,\pi_k)=-\frac{t_P}{\hslash}\frac{\pi_k^4}{3}\psi^0,
\end{equation}
where $F$ indicates a source term for the $\mu^2$-order equation that results to be dependent on the zero-order solution.

Now, the zero-order PDE \eqref{TDSEzeroorder} is the Schr\"odinger equation of a time-dependent harmonic oscillator with standard operators, but in the momentum polarization; therefore the solution $\psi^0(\eta,\pi_k)$ is just the Fourier transform of $\psi_n(\eta,\xi_k)$. In order to derive it, we first rewrite the $\xi_k$ solution \eqref{psin} as
\begin{equation}
    \psi_n(\eta,\xi_k)=\frac{h_n(\frac{\xi_k}{\sqrt{\hslash\,}\,f})}{\sqrt{2^n\,n!\,}}\,\frac{e^{-\frac{R\,\xi_k^2}{2\hslash f^2}}}{(\pi\hslash f^2)^\frac{1}{4}}\,e^{i\alpha_n},
    \label{newpsinxi}
\end{equation}
where $R$ has been defined in \eqref{defineR} and depends on $\eta$. Even though this depends on time through $f$, $R$ and $\alpha_n$, this dependence doesn't affect the implementation of a Fourier transform. Indeed, we can define
\begin{equation}
    \psi_n(\eta,\pi_k)=\int_{-\infty}^{\infty}\frac{d\xi_k}{\sqrt{2\pi\hslash}}\,\psi_n(\eta,\xi_k)\,e^{-i\,\frac{\xi_k\pi_k}{\hslash}},
    \label{fouriertransform}
\end{equation}
\begin{equation}
    \psi_n(\eta,\xi_k)=\int_{-\infty}^{\infty}\frac{d\pi_k}{\sqrt{2\pi\hslash}}\,\psi_n(\eta,\pi_k)\,e^{i\,\frac{\xi_k\pi_k}{\hslash}},
\end{equation}
and insert the last expression inside equation \eqref{timedependentSchroedingerxi}; we see that for the left-hand side, the time derivative can go inside the integral and it affects only $\psi(\eta,\pi_k)$; regarding the right-hand side, the second derivative after $\xi_k$ can also enter the integral, and this time it affects only the exponential, yielding $-\pi_k^2/\hslash^2$, while for the $\xi_k^2$ term (that can also go inside the integral) we rewrite it as a second derivative after $\pi_k$ of the exponential and then we need to integrate by parts twice in order to obtain the term ${\hslash^2\partial^2\psi/\partial\pi_k^2}$. Therefore we see that the Fourier transform of the solution of equation \eqref{timedependentSchroedingerxi} satisfies equation \eqref{TDSEzeroorder}. Now, the expression \eqref{newpsinxi} for $\psi_n(\eta,\xi_k)$ is just a slightly more complicated version of a Gaussian times a Hermite polynomial, so we can already suppose that its Fourier transform will have a similar form; indeed, by computing the integral \eqref{fouriertransform}, we find
\begin{equation}
    \hspace{-0.3cm}\psi_n^0=(-i)^n\left(\frac{R^*}{R}\right)^\frac{n}{2}h_n\bigl(\frac{\pi_k\,f}{\sqrt{\hslash\,}\,\abs{R}}\bigr)\sqrt{\frac{f}{2^n\,n!\,R\,\sqrt{\pi\hslash\,}\,}\,}\,e^{-\frac{\pi_k^2f^2}{2R\hslash}}e^{i\alpha_n},
\end{equation}
which is again a Hermite polynomial times a Gaussian with inverted variance; the phase term containing $\alpha_n$ depends only on time and is thus unaffected, the term $(R^*/R)^{n/2}$ normalizes the Hermite polynomials, and the factor $(-i)^n$ is needed to make the action of the ladder operators consistent. This expression is a solution of the momentum-space Schr\"odinger equation \eqref{TDSEzeroorder}, is normalized and satisfies all the needed relations; besides, it again reduces to the standard momentum-space solution of the time-independent harmonic oscillator under the substitution $f\to1/\sqrt{\omega_k}=$ const.

Now, looking at the first order PDE \eqref{TDSEfirstorder}, it is the same of the zero order one but with the addition of the source term $F(\eta,\pi_k)$. In order to solve it, we consider that the eigenfunctions $\psi_n^0(\eta,\pi_k)$ form a complete orthonormal basis such that $\braket{\psi_{n_1}}{\psi_{n_2}}=\delta_{n_1,n_2}$ and any function can be expressed as a linear combination of them. Therefore we can write $\psi^1$ and $\psi^0$ as
\begin{equation}
    \psi^0=\sum_n\,c_n(\eta)\,\psi_n(\eta,\pi_k),\quad\psi^1=\sum_n\,d_n(\eta)\,\psi_n(\eta,\pi_k),
\end{equation}
where $c_n(\eta)$, $d_n(\eta)$ are time-dependent coefficients; when we plug these expansions back into the first order Schr\"odinger equation \eqref{TDSEfirstorder} we are left with just a recurrence relation for the coefficients, since all the eigenfunctions $\psi_n^0$ satisfy the zero-order equation \eqref{TDSEzeroorder} that corresponds to the homogeneous part of the first order one:
\begin{equation}
    i\hslash\sum_n\,\dv{\,d_n}{\eta}\,\psi_n^0(\eta,\pi_k)=-\frac{t_P}{\hslash}\,\frac{\pi_k^4}{3}\,\sum_n\,c_n(\eta)\,\psi_n^0(\eta,\pi_k).
    \label{recurrence}
\end{equation}
Considering just the ground state and using the result \eqref{resultofpik} for $\pi_k$, we obtain
\begin{equation}
    \hspace{-0.4cm}\pi_k^4\,\psi_0=\frac{3\hslash^2}{4}\frac{R^2(R^*)^2}{f^4}\psi_0-\frac{3\hslash^2}{\sqrt{2\,}}\frac{R^3R^*}{f^4}e^{2i\varphi}\psi_2+\sqrt{\frac{3}{2}\,}\,\frac{\hslash^2R^4}{f^4}e^{4i\varphi}\psi_4;
\end{equation}
it is thus clear that, when $c_n=\delta_{0,n}$, the only non-zero coefficients on the left hand side are $d_0$, $d_2$ and $d_4$. Therefore the relations for these coefficients are:
\begin{equation}
    i\dv{\,d_0}{\eta}=-\frac{t_P}{4}\,\frac{(1+f^2f'^2)^2}{f^4},
    \label{d0}
\end{equation}
\begin{equation}
    i\dv{\,d_2}{\eta}=+\frac{t_P}{\sqrt{2\,}}\,\frac{(1+f^2f'^2)(1-iff')^2}{f^4}e^{2i\varphi},
    \label{d2}
\end{equation}
\begin{equation}
    i\dv{\,d_4}{\eta}=-\frac{t_P}{\sqrt{6\,}}\,\frac{(1-iff')^4}{f^4}e^{4i\varphi}.
\end{equation}
Finally, the ground state of our system in the $\pi_k$ representation is
\begin{equation}
    \psi_0^\text{tot}(\eta,\pi_k)=\psi_0^0(\eta,\pi_k)+\mu^2\,\sum_{n=0}^2d_{2n}(\eta)\,\psi_{2n}^0(\eta,\pi_k).
\end{equation}
From here on we will omit the superscript indicating the order, since we expressed $\psi^1$ and $\psi^0$ as linear combinations of $\psi_n(\eta,\pi_k)$.

Now, in order to find the final spectrum of perturbations we have to evaluate the expectation value $\ev{\hat{\xi}_k^2}$ on the ground state; we can use the expression \eqref{resultofxik} and therefore write
\begin{equation}
\begin{aligned}
    &\ev{\hat{\xi}_k^2}{\psi_0^\text{tot}}=\int d\pi_k\,{\psi_0^\text{tot}}^*\hat{\xi}_k^2\psi_0^\text{tot}=\int d\pi_k\abs{\hat{\xi}_k\psi_0^\text{tot}}^2=\\
    &=\int d\pi_k\,\hslash\,f^2\abs{\frac{1+\mu^2d_0}{\sqrt{2\,}}\,e^{i\varphi}\psi_1+\mu^2d_2e^{-i\varphi}\psi_1+...}^2,
\end{aligned}
\end{equation}
where the dots stand for terms proportional to $\psi_3$ and $\psi_5$, whose square modulus would contribute with terms of order $\mu^4$ which we would neglect. The norm of $\psi_0^\text{tot}$ is easily calculated to be $\abs{N}^2=1+2\mu^2\Re(d_0)$, since $\int d\pi_k\abs{\psi_n}^2=1$, and thus the normalized expectation value of $\hat{\xi}_k^2$ results to be
\begin{equation}
\begin{aligned}
   \frac{\ev{\hat{\xi}_k^2}}{\,\abs{N}^2}&=\frac{\hslash\,f^2}{2\abs{N}^2}\left(1+2\mu^2\Re(d_0)+2\sqrt{2\,}\,\mu^2\Re(d_2e^{-2i\varphi})\right)=\\
    &=\frac{\hslash\,f^2}{2}\,\left(1+\frac{2\sqrt{2\,}\,\mu^2\Re(d_2e^{-2i\varphi})}{1+2\mu^2\Re(d_0)}\right).
\end{aligned}
\end{equation}
As expected, the zero-order term is the same as for the standard Spectrum \eqref{csiquadrostandard}; on the other hand, for the $\mu^2$-order correction we see that we only need $d_0$ and $d_2$ among the coefficients of the expansion.

\begin{figure}
    \centering
    \includegraphics[width=0.9\linewidth]{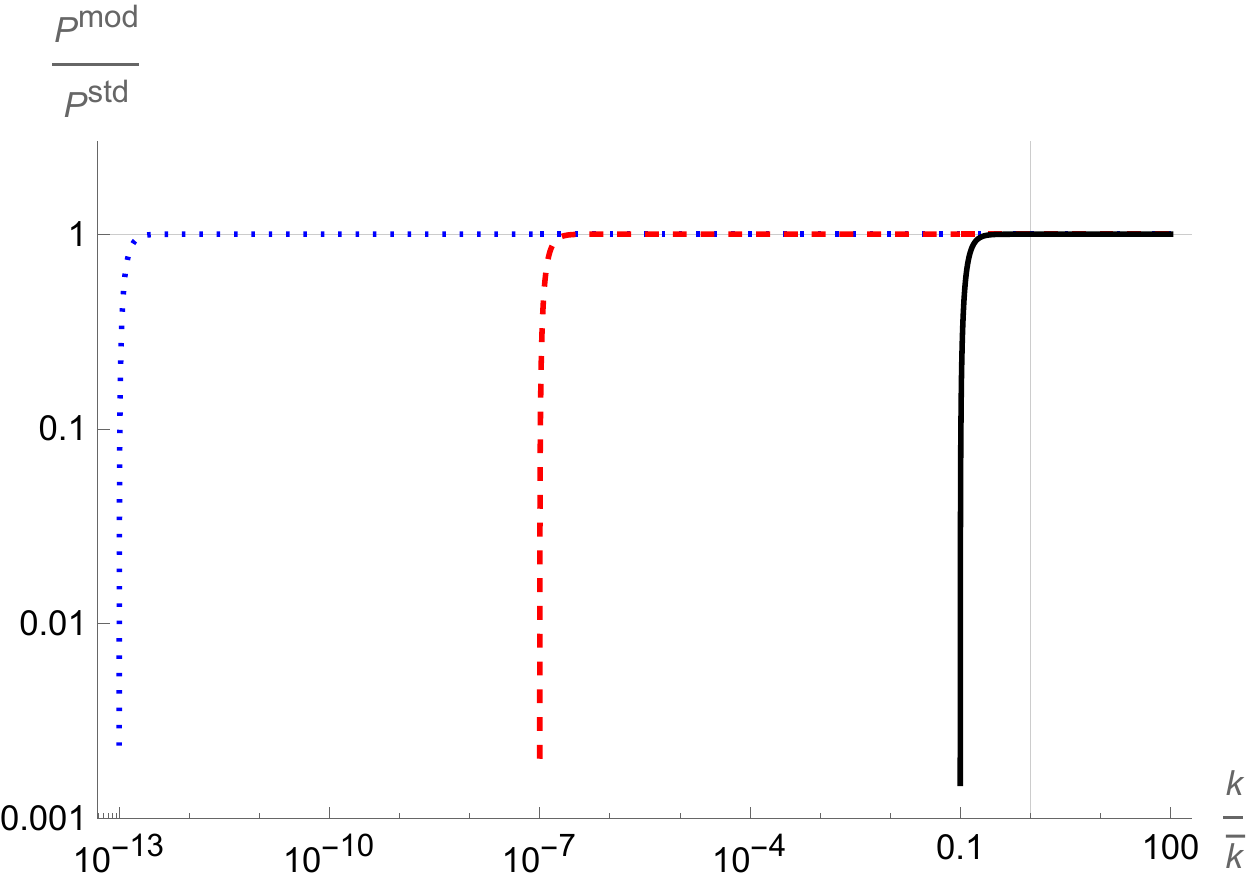}
    \caption{The modified Power Spectrum $\mathcal{P}^\text{mod}$ rescaled to the standard one $\mathcal{P}^\text{std}$ for $\mu=10^{30}$ (black continuous line), $\mu=10^{15}$ (red dashed line) and $\mu=1$ (blue dotted line). The pivotal scale $k=\overline{k}$ and the standard flat spectrum are indicated by faded grey lines.}
    \label{spectrum}
\end{figure}

Now, looking at equation \eqref{d0} we see that the right hand side is real; therefore $d_0$ has a purely imaginary time derivative, and its real time-independent part must be set through initial conditions; we will adopt the same prescription as in \cite{BrandenbergerInit,polymatter} where we assume that the wavefunction is in the instantaneous ground state at the beginning of inflation: we therefore write $d_0(\eta_s)=0$ and, since its real part is independent of time, it will remain zero throughout the evolution. Then we solve the integral \eqref{d2} for $d_2(\eta)$, insert it into the Spectrum and find the asymptotic behaviour:
\begin{equation}
\begin{aligned}
    \mathcal{P}^\text{mod}(k)=&\eval{\frac{c^2k^3}{4\pi^2}\,\frac{\hslash f^2}{2a^2\epsilon}\left(1-2\sqrt{2\,}\,\mu^2\Re(d_2e^{-2i\varphi})\right)}_{-ck\eta\ll1}=\\
    &=\eval{\frac{\hslash}{c}\,\frac{H_s^2}{8\pi^2\epsilon}\left(1-\frac{4t_P\mu^2}{7c^5k^5\eta^6}\right)}_{-ck\eta\ll1}.
\end{aligned}
\end{equation}
Now, if we performed the limit $-ck\eta\to0$ our correction would diverge; however, inflation doesn't actually go on forever but ends at some finite instant; therefore we choose to compute the Spectrum at the value $\eta=\eta_f$ that is the end of inflation. Then we can set ${\eta_f=2\pi/c\overline{k}}$ where $\overline{k}$ is a pivot scale and, by choosing the standard pivotal scale $\overline{k}=0.002Mpc^{-1}$ used in the analysis of the CMB spectra, the spectrum can then be rewritten as
\begin{equation}
\begin{aligned}
    \mathcal{P}^\text{mod}(k)&=\frac{\hslash}{c}\,\frac{H_s^2}{8\pi^2\epsilon}\left(1-\frac{4}{7}\frac{ct_P\overline{k}}{(2\pi)^6}\,\mu^2\,\biggl(\frac{\overline{k}}{k}\biggr)^5\right)=\\
    &\approx\mathcal{P}^\text{std}\left(1-10^{-65}\,\mu^2\,\biggl(\frac{\overline{k}}{k}\biggr)^5\right);
\end{aligned}
\end{equation}
then, by asking that at the pivot scale $k=\overline{k}$ corrections be of order lower than $10^{-3}$, we obtain a constraint on the deformation parameter:
\begin{equation}
    \mu<10^{31}.
\end{equation}
In Figure \ref{spectrum} we see the modified Power Spectrum (rescaled to the standard one) for different values of the deformation parameter $\mu$; the result is a suppression of the Spectrum for small values of $k$ (corresponding to large scales), and the magnitude of the suppression depends on the deformation parameter $\mu$.

We conclude by noting that our construction is similar to the implementation of modified dispersion relations, with the exception that those would be implemented as having a different function $\omega_k(k)$, i.e. they would affect the form of the auxiliary equation \eqref{auxiliary}, while our modifications are implemented at a more fundamental level on the commutation relations. Different forms of modified dispersion relations have been analyzed in the past, but they usually predict a red tilt of the Spectrum (either in the form of a suppression at high energies or of an infrared divergence), exotic behaviours such as oscillations in certain ranges, or no correction at all (see for example \cite{DispersionRelations1,DispersionRelations2,DispersionRelations3,DispersionRelations4}). On the other hand, computations of the Primordial Power Spectrum in Emergent Universe models obtained with different mechanisms have been shown to sometimes yield a suppression at large scale similar to ours (although with different magnitude and features) \cite{EUPowerSpectrum1,EUPowerSpectrum2,EUPowerSpectrum3}. Therefore this might perhaps be a general prediction of the kind of constructions that allow for an Einstein-static beginning of the Universe.

\section{Concluding Remarks}
\label{concl}
We started from the so-called Emergent Universe, i.e. a non-singular standard cosmology with positive curvature on which a Cauchy problem is assigned which balances the matter contribution with the spatial curvature of the model. As a result, the initial phases of the cosmological dynamics are characterized by a a non-zero space volume, approached for the synchronous time going to negative infinity.

The main point in the analysis above was the possibility to construct a non-singular dynamics similar to that of an EU model by implementing a modified Uncertainty Principle. The restated symplectic algebra provided in the quasi-classical limit is, de facto, inspired from Polymer Quantum Mechanics when the basic commutation relation is expanded for a small lattice step. We have shown in detail how the picture of an EU properly arises from the implementation of our dynamical scheme to the positively-curved isotropic Universe. There have been other attempts to generate an Emergent Universe scenario, but they usually have specific requirements, such as specific shapes of the potential, the presence of exotic matter or a modified continuity equation for matter (for example, see \cite{EULoop,EUexotic,EUEMSG} and references therein). The relevance of our restated cosmological dynamics consisted in the possibility to have a finite volume limit in the distant past of our Universe without any fine-tuning of the initial conditions or any need for strange forms of matter, but just as a natural and general feature of the modified symplectic algebra, phenomenologically similar to a modified gravity approach. We discussed in detail the different Universe phases in the proposed scheme, with particular emphasis on the possibility to have an inflationary de Sitter period that is well reconnected to the subsequent radiation dominated era, where most part of the actual Universe morphology is determined via Baryogenesis, Nucleosynthesis, and structure formation \cite{Weinberg}.

A relevant part of the proposed study concerned the implementation of the modified Uncertainty Principle to the pure quantum dynamics of the inflaton field. We constructed the Hamiltonian for each Fourier mode of the quantum scalar field, which corresponded to that of a time-dependent harmonic oscillator (as in the standard spectrum case) plus a small perturbation controlled by the value of the cut-off parameter. We then performed a suitable perturbation theory procedure to calculate the modified expectation value of the squared Fourier harmonics of the Mukhanov-Sasaki variable constructed from the inflaton field, hence computing the corrections to the primordial perturbation Spectrum. Finally, we carefully analyzed the constraints and the proper initial conditions we have to impose on our model in order for the correction to the standard spectrum to live in an observational window for future experiments on the microwave background temperature distribution \cite{Weinberg}.

The present model has the merit to make the non-singular EU model a general feature of the isotropic Universe when a specific sector of cut-off physics is addressed. Furthermore, such a non purely classical feature of the Universe dynamics is expected to leave a specific trace on the primordial Spectrum, which could in principle be identified as a fingerprint on the temperature distribution of the microwave background.

It remains as a future investigation objective to determine how general the proposed scenario is, for example by considering more general cosmological models like the Bianchi Universes \cite{MontaniReview}. Clearly, these cosmological frameworks can be reconciled to the isotropic late Universe by the inflationary de Sitter phase \cite{Kirillov2002}, whose associated Spectrum should have a corrected morphology which is expected to be similar to the one presented here.

\section*{Acknowledgements}
G. B. would like to thank L. Falorsi and S. Segreto for useful insight and discussions. G. B. and A.M. also thank the TAsP Iniziativa Specifica of INFN for their support.

\bibliographystyle{apsrev4-2}
\bibliography{EU.bib}

\end{document}